\documentstyle[prl,aps,multicol]{revtex} 
\input epsf 
\newcommand{\E}{{\rm e}} 
\newcommand{\D}{{\rm d}}

\begin{document} 
\draft 
\title{Persistent current in metals with a large dephasing rate} 
\author{P. Schwab}
\address{  Institut f\"ur Physik, Universit\"at Augsburg,
D-86135 Augsburg
} 
\date{\today} 
\maketitle 
\begin{abstract} 
In a weakly disordered metal electron interactions are responsible for both
decoherence of the quasi-particles as well as for quantum corrections to thermodynamic
properties. 
We consider electrons which are interacting with two-level-systems.
We show that the two-level-systems enhance the average equilibrium 
(``persistent'') current
in an ensemble of mesoscopic rings.
The result 
supports the recent suggestion
that two puzzles in mesoscopic physics may be related: The low temperature saturation
of the dephasing time and the high persistent current in rings.
\end{abstract} 
 
\begin{multicols}{2} 
Quantum interference effects play a crucial role in the low temperature properties of
normal metals. Prominent examples are weak localization and the associated 
magnetoresistance. 
Recently it was suggested\cite{mohanty99,kravtsov99} that two of the unresolved problems
in the physics of mesoscopic metals may have a common solution:
The large value of the persistent current in mesoscopic rings and
the low temperature saturation of the dephasing rate which is seen in magnetoresistance
measurements.

The first problem is the large value of the persistent current
in rings. 
L\'evy {\em et al}\cite{levy90} measured the nonlinear response to a magnetic field
of an ensemble of $10^7$ mesoscopic copper rings.
The measured signal corresponds to a current
$I \approx  I_0\sin( 2 \pi \phi/\phi_0) $ circulating in each ring. $\phi $ is the magnetic flux 
which penetrates each ring and $\phi_0 = h/e$ is the flux quantum.
For temperatures in the mK regime the amplitude was 
$|I_0| \approx 0.3$nA per ring,
which is of the order of one elementary charge in the time
$\tau_D$
an electron needs to diffuse around the ring,
$|I_0| \approx 0.6 e/\tau_D =0.6 eE_c/\hbar$.
Here $E_c= \hbar/\tau_D=\hbar D/L^2$ is the Thouless energy,
$D$ is the electron diffusion constant, and $L$ is the circumference of the ring.
Similar results were reported in Refs.\cite{mohanty99,reulet95}.

Theory, when neglecting electron interactions,
predicts a current that is of the order 
$I \sim e \delta /\hbar$, where $\delta $ is the average distance of single
particle levels at the Fermi energy\cite{schmid91,oppen91,altshuler91}. 
With the parameters of the experiment\cite{levy90}, 
$\delta/k\approx 0.2 $mK and $E_c/k \approx 25 $mK, 
the current obtained is about two orders of magnitude too small.
Electron interactions first seemed to improve the situation\cite{ambegaokar90}. 
For Coulomb interaction
it was found that
$I \sim e \mu^* E_c /\hbar$, where $\mu^*$ is a dimensionless number
that characterizes the strength of the 
interaction in the Cooper channel. However estimates of
$\mu^*$ gave a value which is an order 
of magnitude too small when comparing it with the experiment\cite{ambegaokar90}.
Surprisingly an enhancement of the current was also reported in 
presence of a moderate concentration of magnetic impurities\cite{schwab97}, with
$I\sim e (E_c/\hbar) \cdot  \min(\hbar/\tau_s , E_c )/k T $, where $1/\tau_s$ is
the spin-flip scattering rate. This can become larger than the current 
coming from the
Coulomb interaction, 
however since the temperature dependence is different from the
one observed this mechanism has not been considered as a possible explanation
of the experiment in Ref.\cite{levy90}.

The second problem concerns the phase coherence of the electrons.
Whereas it is expected that the dephasing rate goes to zero in the zero
temperature limit\cite{altshuler82} many experiments show a saturation 
at low temperature. Usually this saturation is attributed
to the presence of magnetic impurities or to heating.
However, recently a saturation of the dephasing time
has been observed, also after excluding
these possibilities\cite{mohanty97,gougam99}.
Several attempts have been made to explain the low temperature 
saturation of the dephasing time \cite{golubev98,altshuler98,imry99,zawadowski99,raimondi99}.
It has been argued by Altshuler {\em et al}\cite{altshuler98} 
that non-equilibrium electromagnetic noise can decohere the
electrons without heating them. 
Originally, this non-equilibrium noise was suggested to be due to 
external radiation which couples into the samples. 
On the other hand dephasing could also occur due to internal noise. 
In this case a saturation of the dephasing time could also occur in 
equilibrium.
Experimental evidence is in favor of an internal dephasing mechanism
\cite{mohanty97,gougam99},
however it is open if equilibrium or non-equilibrium processes dominate.
 
Recently Kravtsov and Altshuler\cite{kravtsov99} have extended earlier
work\cite{kravtsov93}  
on the effect of a high frequency electromagnetic field in mesoscopic 
rings
and have shown that non-equilibrium noise leads  
to a directed non-equilibrium current. They then suggested
that both the ``large'' currents 
observed in\cite{mohanty99,levy90,reulet95} and the strong dephasing
are related and non-equilibrium phenomena.

In this paper we demonstrate 
that also for the system in thermal equilibrium
an enhanced persistent current is expected if there is an 
additional electron interaction which gives also rise to strong dephasing.
For the particular model involving two-level-systems (TLS) we find
(1) a diamagnetic current in the low magnetic field limit
(2) a temperature dependence which is close to the experimentally observed
one
(3) an amplitude which depends on the concentration of TLS.  
In the following we first recall some of the theoretical concepts concerning the
persistent currents. We then estimate the persistent current coming from 
TLS and, 
finally, relate the persistent current amplitude and the dephasing rate.

The equilibrium current in a ring which is penetrated
by a magnetic flux $\phi$ is
calculated by taking the derivative of the thermodynamic potential,
$I(\phi) = - {\partial \over \partial \phi} \Omega(\mu,\phi).
$
For simplicity we do not discuss the subtle questions concerning differences between the
canonical $F(N,\phi)$ and the grand canonical thermodynamical potential 
$\Omega(\mu, \phi)$\cite{schmid91,oppen91,altshuler91}.
In an ensemble of weakly disordered rings the disorder configuration will change
from ring to ring, so in order to calculate the average persistent current of an ensemble 
of rings one has to average 
over disorder, 
$I(\phi) \to \langle I(\phi) \rangle_{\rm dis}$.
For non-interacting electrons the 
grand canonical potential,
and therefore the persistent current, 
depends
only exponentially weak on the magnetic flux and one finds only a 
small persistent current\cite{cheung89}.

The situation changes in presence of interactions.
As an example take the Coulomb interaction as in Ref.\cite{ambegaokar90}
and consider the classical expression for the Coulomb energy,
\begin{equation} \label{coulomb}
H = {1\over 2} \int \D {\bf r} \D {\bf r'} v({\bf r} - {\bf r'})
\delta n({\bf r},\phi) \delta n({\bf r'},\phi )
.\end{equation} 
This quantity depends on the magnetic flux $\phi $ even on average since the fluctuations of the
electron density are magnetic flux dependent and may be written as
\begin{equation} \label{eq2}
\langle \delta n ({\bf r},\phi ) \delta n ({\bf r'},\phi ) \rangle_{\rm dis} = 
\sum_m A_m \cos( 4 \pi m \phi/ \phi_0 )  \end{equation} 
with 
\begin{equation} \label{eqam}
A_m = {4 N(\epsilon_F)\over {\cal V}} 
 {\sin^2( k_F |{\bf r - r' }|) \over (k_F |{\bf r - r' }|)^2 } 
kT\sum_{\omega > 0 } \sqrt{\omega \over E_c} \E^{-m \sqrt{\omega/E_c} }
,\end{equation} 
compare Ref.\cite{ambegaokar90,schwab97}. 
Here $\omega = 2\pi n k T$ are bosonic frequencies and ${\cal V}$ is the
volume and $N(\epsilon_F)$ is the density of states of the Fermi level. As a technical remark 
we would mention that Eq.(\ref{eqam}) is obtained by evaluating the diagram with one
particle-particle propagator (cooperon).
The harmonics of the persistent current 
$I(\phi )= \sum_m I_m \sin( 4 \pi \phi m/\phi_0)$ are finally 
found as
$I_m = 16 \mu_0 e/\pi m^2 \tau_D$ with 
$\mu_0 = N(\epsilon_F) \int \D {\bf r} v({\bf r}) \sin^2(k_F r)/(k_F r)^2$. 
Including the exchange energy reduces the current by a factor two, and
higher orders in the interaction reduce the interaction amplitude,
$ \mu_0 \to \mu^* \approx  \mu_0/[1+\mu_0 \ln(\epsilon_F \tau_D)] $.
 
When opening an additional interaction channel one will find an 
additional contribution to the persistent current.
In Ref.\cite{schwab97} this has been demonstrated for magnetic impurities. 
Here we consider the interaction of conduction electrons with nonmagnetic impurities, which we
assume to couple to the electron density. 
The Hamiltonian is of the form
\begin{equation}
\hat  H_{\rm int} =  \int \D x \hat n ({\bf x}) \hat V({\bf x}) 
.\end{equation}
The operator
$\hat V({\bf x})$ that is due to the impurities will be specified more explicitly below.
To second order in this interaction one finds a correction to the free energy 
which is the sum of a Hartree and a Fock like term,
$\delta \Omega = \delta \Omega_{\rm H} + \delta \Omega_{\rm F}$,
which are given by ($\beta = 1/k T $)
\begin{eqnarray}
\delta \Omega_{\rm H} &=& -{1\over 2 } \int_0^{\beta}\D \tau \int \D {\bf x} \int \D {\bf x'}
\langle \hat n ({\bf x} ) \rangle \langle \hat n({\bf x'}) \rangle
\nonumber \\
&&\label{hartree} \times \left[ \langle  \hat V({\bf x}, \tau) \hat V({\bf x'}, 0 ) \rangle 
     - \langle \hat V({\bf x}) \rangle \langle \hat V({\bf x'}) \rangle \right] \\
\label{fock}
\delta \Omega_{\rm F} &=& - {1\over 2 }\sum_{s,s'} \int_0^{\beta}\D \tau \int \D {\bf x} \int \D {\bf x'}
\langle \Psi^\dagger_s({\bf x},\tau ) \Psi_{s'}({\bf x'},0 )\rangle \nonumber\\
&& \times
\langle \Psi_s({\bf x},\tau ) \Psi^\dagger_{s'}({\bf x'},0 ) \rangle 
\langle \hat V (\rm x, \tau) \hat V({\rm x'},0 )\rangle 
,\end{eqnarray}
where $\Psi^\dagger_s({\bf x,\tau} )$ and $\Psi_s({\bf x}, \tau )$ are
operators for fermions with spin $s$
and the brackets
$\langle \dots \rangle$ are the thermal average .
If $\hat V({\bf x})$ describes pure potential scattering, 
then $\hat V({\bf x})$ is a c-number with the result that $\delta \Omega_{\rm H}=0$. 
$\delta \Omega_{\rm F}\ne 0 $ but does not depend on the magnetic flux which can be traced back 
to the fact that $\hat V({\bf x},\tau)= \hat V({\bf x})$ is static.
This can become different if the impurity has an internal degree of freedom.
Consider a TLS, realized by an impurity which sits in a double well
potential with minima at
${\bf r}$ and ${\bf r}+{\bf d}$ which are nearly degenerate in energy. 
We write the scattering potential as
$\hat V({\bf x}) = V [\hat n_A \delta({\bf x} -{\bf r} ) +  
                     \hat n_B \delta({\bf x} -{\bf r - d })]$. 
$\hat n_A$ and $\hat n_B$ are the number operators
for the impurity in the relevant potential minimum. 
Since the impurity is in either of these minima
$\hat n_A + \hat n_B = 1 $. 
We further characterize the impurity by 
the asymmetry $\epsilon$ and a tunneling amplitude $\Delta$
between between the two minima, so the impurity Hamiltonian is 
\begin{equation}
\hat H_{\rm imp} = \left( \begin{array}{cc} 
\epsilon & \Delta  \\
\Delta & -\epsilon \end{array} \right)
.\end{equation}
The Hartree energy (\ref{hartree}), which is nonzero in this model, may be interpreted from 
the point of view of both the electrons and the impurities.
From the electronic point of view the electron impurity interaction gives rise to an 
effective electron-electron interaction:
Comparing Eqs.(\ref{coulomb}) and (\ref{hartree}) 
one realizes that the Coulomb interaction is replaced by an 
effective interaction  
\begin{eqnarray}
v({\bf x} - {\bf x'} )  \to - \int_0^\beta \D\tau 
\left\{ \langle \hat V({\bf x}, \tau) \hat V({\bf x'} ,0) \rangle -
\langle \hat V({\bf x}) \rangle \langle \hat V({\bf x'}) \rangle \right\}
\end{eqnarray}
due to the defects.
From the impurity point of view the coupling to the conduction electrons
changes the level asymmetry,
\begin{equation}
\left( \begin{array}{cc}
\epsilon & \Delta \\ \Delta & -\epsilon \end{array} \right) 
\to \left( \begin{array}{cc}
\epsilon + V \langle \hat n({\bf r})\rangle & \Delta \\ \Delta  & 
-\epsilon + V\langle \hat n({\bf r+d)} \rangle \end{array} \right)
,\end{equation}
which then 
changes the free energy as given in Eq.(\ref{hartree}) to second order in $V$.

We discuss the persistent current first in the most simple situation, 
where we neglect the tunnel splitting $\Delta $.
In this case $\hat V({\bf x}, \tau )$ is static so $\delta \Omega_{\rm F}$ remains
independent of magnetic flux, as in the case of ``normal'' disorder. 
Here and below we will therefore concentrate on the Hartree energy.
Using the relation $\hat n_A+ \hat n_B=1$ and averaging over ``normal'' disorder 
we can rewrite the Hartree energy as
\begin{eqnarray}\label{eq11}
\langle \delta \Omega_{\rm H}\rangle_{\rm dis}& = &- |V|^2 
\langle \delta n^2({\bf r } ,\phi) \rangle_{\rm dis}        
\left( 1- {\sin^2 (k_F d) \over (k_F d )^2 } \right) \nonumber \\
&& \times \int_0^\beta \D \tau  
\left\{ \langle \hat n_A(\tau) \hat n_A(0) \rangle - 
\langle\hat n_A \rangle \langle \hat n_A \rangle \right\}
.\end{eqnarray}
If the TLS asymmetry is large, $| \epsilon | > k T$, then 
$\langle \hat n_A(\tau) \hat n_A(0) \rangle - 
 \langle \hat n_A \rangle \langle \hat n_A \rangle =0 $ and therefore 
$\langle \delta \Omega_{\rm H}\rangle_{\rm dis} =0$. For a TLS with a small asymmetry,
$|\epsilon| < k T$ one finds    
$\langle \hat n_A(\tau) \hat n_A(0) \rangle - 
 \langle \hat n_A \rangle \langle \hat n_A \rangle =1/4 $ so that 
$\langle \delta \Omega_{\rm H}\rangle_{\rm dis} \ne 0$ 
and a persistent current results. 
From the integration over $\tau$ it follows that 
the current coming from a single defect 
is proportional to
the inverse temperature, in full analogy to the persistent current
from a magnetic impurity\cite{schwab97}.
For the system with a finite density
of TLS the asymmetry $\epsilon $ will not be a constant, instead there will be a distribution
of asymmetries.
Using eq.(\ref{eq2}) and below we determine
the persistent current as
\begin{equation} \label{eq12}
I  \approx - {8 \over \pi }{c_{\rm act} N(\epsilon_F) V^2 F \over kT }{e\over \tau_D }
,\end{equation}
where $F = 1- \sin^2(k_F d)/(k_Fd)^2 $ and
$c_{\rm act}$ is the concentration of TLS with $\epsilon < kT$ and therefore
is active in producing a persistent current. 
Assuming a flat distribution of asymmetries between zero and $\epsilon_{\rm max} > k T$,
the concentration of active TLS is proportional to the temperature,
$c_{\rm act} = c kT/\epsilon_{\rm max}$, which then cancels
the inverse temperature dependence of the persistent current of a single defect.
The current is diamagnetic in contrast to the paramagnetic current from the
repulsive Coulomb interaction. 
The amplitude of the current is of the diffusive scale, $I \sim e/\tau_D$, 
as for the Coulomb interaction. The 
dimensionless prefactor $\mu^*$ is replaced by the factor  
$\mu_{\rm TLS}= - c F N(\epsilon_F)V^2/\epsilon_{\rm max}$.
which should be of order one
if this mechanism is relevant for the currents observed in Ref.\cite{levy90}.
Assuming an atomic scattering cross section of the TLS and the factor $F\sim 1 $
this requires 
a density of states of TLS that is comparable to the
density of states of the metallic host and therefore
of the order $10^{18}/{\rm K cm^3}$. At 100mK this corresponds to a concentration of active
two-level-systems of about 2ppm which is not a small number but, in principle not impossible
\cite{ralph92}.
For the assumed distribution of asymmetries
the temperature dependence of the persistent current is only due to the temperature dependence of
the local density fluctuations, see eq.(\ref{eqam}), and is therefore 
identical to the temperature dependence of the persistent current from the Coulomb interaction.
The latter has been shown\cite{ambegaokar90} to agree well with the experiment 
in Ref.\cite{levy90}.
Finally it is important to discuss spin-orbit scattering, since in the gold or copper rings in 
the experiments the spin-orbit rate is large. 
Following Refs.\cite{ambegaokar90,schwab97} we find 
that spin-orbit scattering reduces the persistent current due to
the mechanism discussed here
by a factor four, but the sign remains diamagnetic.

Let us now allow a finite tunnel splitting  $\Delta $, i.e.
spontaneous transitions of the impurity between the two minima.
The correlation function that is relevant for the persistent current, i.e. 
the impurity susceptibility, is 
given by
\begin{equation} \label{eq13}
\int_0^\beta \D \tau
\langle \hat n_A(\tau) \hat n_A(0) \rangle - \langle \hat n_A \rangle \langle \hat n_A \rangle 
= \left\{ \begin{array}{lc}
{1\over 4} {1\over k T } 
& \\ 
{1\over 4} {\Delta^2 \over \epsilon^2 + \Delta^2 }{1\over \sqrt{\epsilon^2 + \Delta^2}}
&   
\end{array}
\right.
,\end{equation}
in the two limits where $\epsilon^2 + \Delta^2 < (kT)^2$ and $\epsilon^2 + \Delta^2 > (kT)^2$.
Whereas for static defects with $\Delta=0$ the correlation function is non-zero only 
in the high temperature limit, $kT > \epsilon$, the correlation function for     
dynamic defects is non-zero even in the zero temperature limit, so these defects
contribute to the persistent current even for $T\to 0$.
We calculate the persistent current under the assumption\cite{black81}  
of a flat distribution of   
$\epsilon$ between zero and $\epsilon_{\rm max}$ and a distribution of $\Delta $ that
is proportional to $1/\Delta $ between $\Delta_{\rm min}$ and $\Delta_{\rm max}$.   
We then find $I \sim -(e/\tau_D) F \hbar /(\tau_{\rm TLS}\epsilon_{\rm max})$ 
as before when we neglected the tunnel splitting. $\hbar/\tau_{\rm TLS} \sim c N(\epsilon_F) V^2$
is the electron scattering rate off the TLS.

Finally we discuss the relation of the persistent current and dephasing.
In Ref.\cite{imry99} it has been demonstrated that TLS
lead to dephasing with a rate that is temperature independent in a
certain range of temperature. 
Both the persistent current amplitude and the dephasing rate are hard to estimate
for a given material since they depend on the concentration of TLS and the distribution
of $\epsilon$ and $\Delta$. It is therefore of interest to relate 
the two quantities, in order to reduce the number of unknown parameters.
Notice that
in order to have dephasing there must be real transitions between two impurity states, and
one finds that the defects with 
$kT> \sqrt{\epsilon^2+\Delta^2} > \hbar/\tau_\phi$ 
are most effective for dephasing.
On the other hand all defects with  
$kT > \sqrt{\epsilon^2+\Delta^2} $ and even some with
$kT < \sqrt{\epsilon^2+\Delta^2}$ contribute to the persistent current,
see Eq.(\ref{eq13}). 
We cannot therefore give a general relation between dephasing rate and persistent current amplitude.
We can, however, as shown below,
give such a relation for our special choice of the distribution of
$\epsilon$ and $\Delta$.
The dephasing rate has been estimated as \cite{imry99} 
\begin{equation}
{1\over \tau_\phi} \sim \left\{ \begin{array}{ll} 
 \Delta_{\rm max}F/(\epsilon_{\rm max} \tau_{\rm TLS} \lambda  )  
& {\rm if}\,\, \hbar/\tau_\phi < \Delta_{\rm max} < k T\\
\Delta_{\rm max} ( F /\hbar \lambda \epsilon_{\rm max} \tau_{\rm TLS} )^{1/2} 
& {\rm if }\,\,
\Delta_{\rm max}< \hbar/\tau_\phi  < k T\end{array}\right. 
\end{equation}
with $\lambda = \ln( \Delta_{\rm max}/ \Delta_{\rm min} ) $.
The persistent current amplitude,  
$I \sim \mu_{\rm TLS} (e/\tau_D)	$ with 
$|\mu_{\rm TLS}|\sim F\hbar/(\epsilon_{\rm max} \tau_{\rm TLS})$,
is therefore of the order 
\begin{equation}
|\mu_{\rm TLS}| \sim \left\{ \begin{array}{l}
 \lambda (\hbar/\tau_\phi)/ \Delta_{\max} \\
 \lambda (\hbar/\tau_\phi)^2/\Delta_{\rm max}^2
\end{array}\right.  
\end{equation}
in the two limits considered.
For example for the gold sample of Ref.\cite{mohanty99} $\hbar/\tau_\phi \sim 2$mK below
$500$mK.
If the constant dephasing rate is from the mechanism we consider, then the lowest measured
temperature ($\sim 40$mK) is an upper limit for $\Delta_{\rm  max}$, and leads to the estimate
$|\mu_{\rm TLS}| > \lambda /20$.  

The dephasing rate for low temperature, $\hbar/\tau_\phi<k T< \Delta_{\rm max}$,
is proportional to $T$\cite{imry99} and given by
$1/\tau_\phi \sim  F kT/(\epsilon_{\rm max} \tau_{\rm TLS} \lambda  )  $. 
Here one finds $| \mu_{\rm TLS} | \sim \lambda (\hbar/\tau_\phi)/kT$, which depends
only on one unknown parameter, $\lambda$. 
A dephasing rate which is linear in $T$ has been observed in various three-dimensional and two-dimensional
samples\cite{Ovadyahu84}. The values which were reported 
for $\tau_\phi$ at $10$K are around $\tau_\phi \sim
10^{-12}$sec -- $5 \cdot 10^{-10} $sec, which corresponds to 
$( \hbar/\tau_\phi )/kT \sim 2\cdot 10^{-2}$ -- $1$. Also from these 
considerations it seems rather reasonable
that the parameter $| \mu_{\rm TLS} |$ can reach values of order one.

In this paper we estimate the persistent current linear in the concentration
of TLS and we neglect Kondo correlations. 
Kondo physics has been suggested as a possible solution of the 
dephasing problem in Ref.\cite{zawadowski99}.
The persistent current, of course, will be modified by Kondo correlations,
however it is beyond the scope of this paper to estimate this quantitatively. 
We also do not attempt to give an exhaustive discussion of
the limit of high concentration of impurities. However it is clear that
our theory will overestimate the current when the phase coherence time 
$\tau_\phi$
becomes of the order of, or shorter than, the diffusive time $\tau_D$.
The related problem for the persistent current coming from magnetic impurities
has been discussed in Ref.\cite{schwab97}.

In summary we demonstrated that the interaction of conduction electrons with 
impurities induces a persistent current.
Under reasonable assumptions we find a temperature dependence 
that is set by the diffusive scale.
The most crucial point however is the current amplitude here given by
$I \sim - |\mu_{\rm TLS}| e/\tau_D$. The dimensionless parameter $\mu_{\rm TLS}$ depends
on the concentration of the TLS, so a reliable estimate of the
current amplitude is difficult.
Experimentally the interrelationship of dephasing and persistent current may be checked
by measuring the persistent current for different materials.
For silver, where no saturation of the dephasing time has been observed\cite{gougam99},
we expect a smaller persistent
current than in gold or copper where the dephasing time saturates at low temperature.  
The sign of the current may help to decide if non-equilibrium fluctuation suggested in
Ref.\cite{kravtsov99} or the equilibrium electron-impurity interactions 
studied here dominate the current: 
For a system with strong spin-orbit interactions Ref.\cite{kravtsov99} predicts a paramagnetic current,
whereas we found a diamagnetic current.

We acknowledge stimulating discussions with U. Eckern
and financial support by the DFG through SFB 484
and Forschergruppe HO/955.

\end{multicols}

\end{document}